\documentclass[12pt]{article}
\PassOptionsToPackage{bookmarksopen,bookmarksnumbered,colorlinks,linkcolor=blue,citecolor=blue,urlcolor=blue}{hyperref}

\usepackage[dvipsnames]{xcolor}
\usepackage{times}
\usepackage[left=3cm, right=3cm, top=2.5cm, bottom=2.5cm]{geometry}
\usepackage{fix-cm}

\usepackage{xcolor, soul}
\sethlcolor{yellow}

\usepackage{fancyhdr}
\usepackage{graphicx}
\usepackage{caption}
\usepackage{subcaption}
\usepackage{float}

\usepackage{array}
\usepackage{tabularx}
\usepackage{longtable}
\usepackage{dcolumn}
\newcolumntype{d}[1]{D{.}{.}{#1}}
\newcolumntype{Y}{>{\centering\arraybackslash}X}
\usepackage{booktabs}
\usepackage{multirow}
\usepackage{colortbl}

\usepackage{amsmath}
\usepackage{bm}

\usepackage{orcidlink}

\usepackage{parskip}
\usepackage{lineno}
\usepackage{lipsum}

\usepackage{natbib}

\usepackage{hyperref}

\newcommand{\doctitle}[1]{{\centering \fontsize{14}{16}\selectfont \textbf{#1} \par}}
\newcommand{\docauthor}[1]{{\centering \fontsize{12}{14}\selectfont \textbf{#1} \par}}
\newcommand{\docaffiliation}[1]{{\centering \fontsize{11}{13}\selectfont #1 \par}}
\newcommand{\dockeywords}[1]{{\fontsize{12}{14}\selectfont \textbf{Keywords :} #1 \par}}

\makeatletter
\renewcommand\section{\@startsection {section}{1}{1em}%
                                   {-3.5ex \@plus -1ex \@minus -.2ex}%
                                   {2.3ex \@plus.2ex}%
                                   {\normalfont\fontsize{12}{14}\bfseries}}

\renewcommand\subsection{\@startsection {subsection}{2}{1em}%
                                   {-3.25ex \@plus -1ex \@minus -.2ex}%
                                   {1.5ex \@plus .2ex}%
                                   {\normalfont\fontsize{12}{14}\bfseries}}

\makeatother

\begin{document}
\setlength{\bibsep}{0.5\baselineskip}
\doctitle{A generalised discrete mixture model to better capture preference heterogeneity in discrete choice data}
\hfill\break%
\docauthor{Thomas O. Hancock\textsuperscript{1,2} \&\
           John Buckell\textsuperscript{2,3}
           }
\docaffiliation{\textsuperscript{1}Choice Modelling Centre, Institute for Transport Studies, University of Leeds, UK\\ 
                \textsuperscript{2}Nuffield Department of Primary Care Health Sciences, University of Oxford, UK\\
                \textsuperscript{3}Nuffield Department of Population Health, University of Oxford, UK}
\hfill\break%
\dockeywords{Discrete choice models; latent class models; discrete mixture models; preference heterogeneity.} 

\section*{Abstract}
Arguably the key issue in modelling discrete choice data is capturing preference heterogeneity. This can be through observed characteristics, and/or using techniques for capturing random heterogeneity across respondents. On the latter, in health economics, the two main approaches are the mixed multinomial logit (MMNL) and the latent class (LC) model. In this paper, we revisit the discrete mixture (DM) model as a third alternative to these. The DM model is similar to LC but allows for any combination of preferences across attributes, rather than grouping preferences as is the case in LC. We next develop a generalised discrete mixture (GDM) model. Additional boosting parameters in the class allocation component allow the model to collapse to a standard DM or LC structure as best fits the data at hand. This means that the model, by definition, performs at least as well as the best of a standard DM and a LC model; or better than both. Additional benefits include that it (a) allows the data to tell us the underlying correlations of preferences, (b) does not rely on distributions as is the case for mixed logit models, meaning estimation times are reduced and it does not require assumptions on the distribution of preferences. Exercises on simulated data show the unlikely conditions under which a LC model would be preferred to a DM. The convention of labelling latent classes, we believe, is questionable in many cases. The GDM is suitable in all cases. We show in empirical data that the GDM substantially outperforms LC models, granting a more detailed depiction of respondents' preferences.  

\section{Introduction}
A key concern in modelling choice behaviour is to account for heterogeneity in decision-making. 
Capturing heterogeneity in decision-making is crucial, as it allows for the development of better choice models to more accurately predict and forecast choice behaviour, whilst simultaneously delivering insights on how different segments of the population will react in a given context.  Recognition of the fact that decision-makers face complex choice tasks has led to the development of many different methods to account for various sources of heterogeneity in decision-making, including (1) \textit{Taste} heterogeneity, where the relative impact of an attribute on the different alternatives being chosen can vary across different decision-makers, i.e. some individuals may care about a given attribute (e.g. cost) much more than other individuals do. (2) \textit{Information processing} heterogeneity, where the amount of information considered by different decision-makers may vary. For example, some decision-makers may use all attributes to make their decision, while others may exhibit \textit{attribute non-attendance} and not use all available information (see for example \citealp{hensher2009simplifying,erdem2015accounting,heidenreich2018decision}).  (3) \textit{Decision-rule} heterogeneity, where it is assumed that individuals use different deliberative mechanisms to make decisions, i.e. some individuals may maximise their utility, while others may minimise regret \citep{chorus2008random,chorus2010new}. Recognising the different forms of heterogeneity and accounting for these in modelling may lead to a better explanatory power of models \citep{buckell2022utility}. However, disentangling different confounding sources of heterogeneity, such as whether an individual does not attend to an attribute or does not care about it, is challenging \citep{hess2013s,hancock2021really}. In the work in this paper, we focus on better accounting for taste heterogeneity, with the important result that doing so may also gives insights on other types of heterogeneity, including specifically information processing heterogeneity. 

\begin{figure}[ht]
    \centering
    \includegraphics[width=0.5\linewidth]{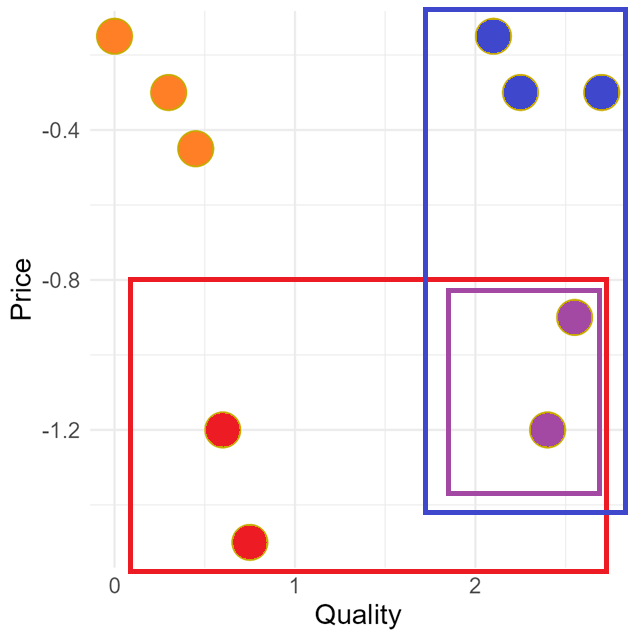}
    \caption{Some individuals may be \textcolor{red}{cost sensitive}. Others may care more about the relative \textcolor{blue}{quality} of different alternatives. Some individuals may be \textcolor{red}{cost} and \textcolor{blue}{quality} sensitive (\textcolor{Plum}{i.e., both}), whilst others might not care much about \textcolor{orange}{either}.}
    \label{fig:PrefVisual}
\end{figure}

Taste heterogeneity, or \textit{preference} heterogeneity, is the most studied form, unsurprisingly as it can give insights on \textit{who} cares about \textit{what}. Typically, there are two key modelling methods to account for differences in preferences across individuals. The first is the use of sociodemographic variables to capture \textit{deterministic heterogeneity}, i.e. differences in preferences that are observable to the analyst, that are based on characteristics of the different decision-makers. The second is the use of different model structures to capture \textit{random heterogeneity}, i.e. typically the use of random parameters in mixed multinomial logit models (MMNL) or latent class (LC) structures to capture unobserved differences in preferences. An example of different possible underlying preferences across 10 individuals is visualised in Figure \ref{fig:PrefVisual}.

In health economics, \citet{Vass2022} and \citet{soekhai2019discrete} document that the vast majority of models allowing for random taste heterogeneity use either MMNL or LC. However, both of these models have distinct issues. For MMNL, whilst recent work has demonstrated the flexibility of MMNLs when allowing for different distributions \citep{buckell2025} and different preferences (e.g. price and quality sensitivities) can be correlated \citep{hess2017correlation}, the simulation of these models is time-consuming, especially if the recommended number of draws when simulating these distributions are used \citep{czajkowski2019simulation}. This is a particularly crucial concern in the era of big data, where databases may contain choice data for thousands or even millions of individuals. 
Latent class models, meanwhile, do not rely on draws, thus can be estimated quickly. They instead specifically assume that each individual is probabilistically assigned to different possible `classes' of preferences. However, first, by definition, preferences for the different attributes are correlated. This is a result of the preferences specifically being grouped within classes. For example, class 1 might be cost insensitive and quality sensitive, whilst class 2 might be cost sensitive and quality insensitive. This does not allow for the possibility of an individual being both cost and quality sensitive. Additional classes within the model structure might work for case studies with few attributes, but in most contexts within health economics, there are a large number of attributes, meaning all possible combinations of preferences are not possible under a latent class construct.   Second, if differences in sensitivity to one particular attribute are key, this will drive the parameter estimates. As a result, efforts to capture other forms of heterogeneity will fail. For example, in the case of choosing between different treatments, some individual may not care about side effects. However, if the key source of heterogeneity is a difference in cost sensitivity, then a latent class model with two classes will only capture insensitivity to side effects if it is perfectly correlated with cost sensitivity. If there is a group of individuals who care about side effects who are more likely in the cost sensitive class, and another group in the cost insensitive class, then both classes will appear to care about side effects, ruling out the possibility of uncovering `non-attendance' of side effects. Third, latent class models are often misinterpreted. They specifically \textit{probabilistically} assign individuals into different classes. Labels for the different classes are frequently based on characteristics of the individuals that are more likely (but not guaranteed) to be in those classes. This erroneously omits the probabilistic element of the model, with, for example, posterior analysis \citep{hess2024latent} rarely applied to bypass this concern. 

We address some of the issues above through the use of a \textit{`discrete mixture'} model based on the specification of \citet{hess2007systematic}. This model estimates different probabilities for each individual having each particular taste, as opposed to a single probability for a set of tastes. Consequently, there are effectively latent classes for different sensitivities to each attribute, and the `discrete mixtures' are the possible combinations of the different preferences (e.g. high or low cost sensitivity and also high or low quality sensitivity). Consequently, standard forms of discrete mixture models assume zero correlation between the different attribute sensitivities. This of course is also problematic, as in the example in Figure \ref{fig:PrefVisual}, it might be expected that more individuals have preferences represented by the dots in the off-diagonals quadrants, i.e. decision-makers tend to prioritise one particular attribute (quality or cost). 

We next develop \textit{`generalised discrete mixture'} models. These models through grouping of attributes or through the class allocation function, allow for correlation of preferences. The latter methods estimates additional boosting parameters that increase the probability of particular combinations of preferences co-occuring. High estimates for these parameters allow the model to collapse towards a latent class structure, where specific combinations of attribute sensitivities always co-occur. Correspondingly, insignificant estimates for these parameters result in the model collapsing towards the typical formulation of the discrete mixture model.

The rest of this paper is organised as follows. First, we briefly give the mathematical formulation of latent class models, discrete mixture models and the new generalised discrete mixture models. Next, we provide visualisation of these different models to a small-scale simulated dataset (based on Figure \ref{fig:PrefVisual}) to highlight how the different models operate. We subsequently present simulated datasets where the correlation between attribute preferences vary, to test the different models. We next apply the models to two different case studies on smoking preference choice experiments. Finally, we conclude by providing 
a discussion on the implications of the findings and summarising ideas for future research.

\section{Mathematical overview of models}

The mathematical formulation of our base multinomial logit model (MNL), latent class model (LC), discrete mixture model (DM) and generalised discrete mixture models (GDM) are given below.

\subsection{Multinomial logit}

Following standard practice, we assume that individual $n$, in choice scenario $s$, derives a utility ($U_{nsi}$) for alternative $i$:

\begin{equation}
    U_{nsi} = V_{nsi} + \varepsilon_{nsi},
\end{equation}

\noindent where $V_{nsi}$ is the utility that is observed by the analyst and $\varepsilon_{nsi}$ is the unobserved utility. The use of a type I extreme value error term, distributed randomly across individuals, choice tasks, and alternatives, gives the standard form of the multinomial logit model. Thus, the probability of alternative $j$ being chosen by individual $n$ in choice task $s$ ($J^*_{ns}$) from from a set of $J_{ns}$ alternatives is calculated:

\begin{equation}\label{eq:MNL}
    P(J^*_{ns}) = P(j|n,s) = \frac{\exp(V_{nsj})}{\sum_{i=1}^{J_{ns}}(\exp(V_{nsi}))}
\end{equation}

\noindent A set of MNL model parameters, $\Theta_{MNL}$, are then optimised to maximise the total log-likelihood of observing all choices ($C$) in a given dataset, such that: 

\begin{equation}\label{eq:LL_MNL}
    LL_{MNL}(C|\Theta_{MNL}) = \sum_{n=1}^{N} \ln\left(\prod_{s=1}^{S_n}  P(J^*_{ns})\right),
\end{equation}

\noindent where each individual $n$ makes a total of $S_n$ choices, and there are $N$ individuals in total.

\subsection{Latent class models}
In a latent class model, it is assumed that the utility for each alternative varies across a set of different `classes' of behaviour. Thus, the preference of an individual will depend upon which classes ($m$) they are probabilistically assigned into. We thus now have $U_{nsi|m}$. Consequently, the log-likelihood for the latent class model is:

\begin{equation}\label{eq:LL_LC}
        LL_{LC}(C|\Theta_{LC}) = \sum_{n=1}^{N} \ln\left(\textcolor{red}{\sum_{m=1}^{M}\pi_{m,n}}\prod_{s=1}^{S_n}  P(J^*_{ns\textcolor{red}{|m}})\right),
\end{equation}

\noindent  where the red text indicates the difference between Eq. \ref{eq:LL_MNL} for MNL and Eq. \ref{eq:LL_LC} for LC. We now instead have $P(J^*_{ns|m})$ which is the probability of the alternative under class $m$, where it should be noted that the functional form of $P(J^*_{ns|m})$ need not follow Eq. \ref{eq:MNL}, i.e. different model structures could be used in the different classes (see \citet{meester2023can} for example).

Individuals $n$ are assigned into the different classes depending on some function (typically incorporating socioeconomic variables, see below), meaning that we have a set of contributing shares for class $m$ for individual $n$, labelled $\pi_{n,1}$ ... $\pi_{n,M}$,  where we have a total of $M$ classes, $\pi_{n,m} \ge 0, \forall m$, and $\sum_{m=1}^{M} \pi_{n,m}=1, \forall n$. The model estimates a class share $\pi_{m^*,n}$ for a given class $m^*$ by calculating:

\begin{equation}\label{eq:LC_socios}
    \pi_{m^*,n} = \frac{\exp\left(\Delta_{m^*} + \sum_{l=1}^{L}(z_{n,l} \cdot \zeta_{l,m^*})\right)}{\sum_{m=1}^{M} \exp\left(\Delta_m + \sum_{l=1}^{L}(z_{n,l} \cdot \zeta_{l,m})\right)},
\end{equation}

\noindent where there are $L$ sociodemographic variables ($z_{n,l}$) describing individual $n$, with corresponding estimated coefficients $\zeta_{l,m}$. There are also class specific constants $\Delta_{m^*}$ such that in the absence of sociodemographics, all individuals will be assigned equivalently across classes ($\pi_{m,n} = \pi_m. \forall n$), with shares thus dependent only on $\Delta_{m^*}$.

\subsection{Discrete mixture models}
A discrete mixture model (DM) has exactly the same model structure as a latent class model (LC), i.e. the likelihood of the model is based on a weighted sum across different classes where each class represents a different set of preferences \citep{hess2007systematic}, such that we have:

\begin{equation}
    LL_{DM} \left( C|\Theta_{DM} \right) = \sum_{n=1}^{N} \ln\left(\sum_{m=1}^{M}\pi_{m,n}\prod_{s=1}^{S_n}  P(J^*_{ns|m})\right),
\end{equation}

\noindent The difference between LC and DM is the number of classes $M$. Whereas a latent class model assumes classes with their own unique set of preferences for each attribute, a discrete mixture assumes that combinations of preferences (\textit{`mixtures'}) are possible. Thus, for example, class 1 of a latent class model might have tastes $\beta_{A_1}$ and $\beta_{B_1}$ for attributes $A$ and $B$ respectively, and class 2 might have tastes $\beta_{A_2}$ and $\beta_{B_2}$. A discrete mixture model would have these two classes (mixtures) \textit{and additionally} mixture 3 ($\beta_{A_1}$ and $\beta_{B_2}$) and mixture 4 ($\beta_{A_2}$ and $\beta_{B_1}$).  

\begin{figure}[ht!]
    \centering
    \includegraphics[scale=0.8]{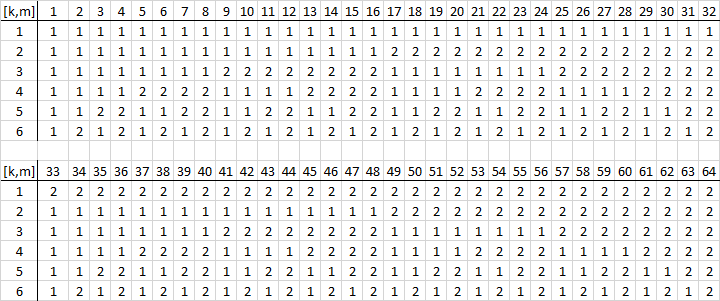}
    \caption{The matrix $\Lambda_{k,m}$, which gives all possible combinations of coefficients, for a case where there are $K=6$ attributes, each with $2$ estimated coefficients. A standard latent class model would not have combinations of coefficients, but would instead simply have two classes (corresponding to $m=1$ and $m=64$, such that there are no shared coefficients between the two classes).}
    \label{fig:combinations}
\end{figure}

The calculation of the shares of the mixtures in DM are thus much more complex than the calculation of the shares of classes in LC. They are defined:

\begin{equation}\label{eq:DM_shares}
    \pi_{m,n} = \prod_{k=1}^{K} \left( \sum_{q=1}^{Q_k} \omega_{n,k,q} \cdot (\Lambda_{k,m} = q) \right),
\end{equation}

where $k = 1,$...$,K$ is an index over attributes, $q = 1,$...$, Q_k$ is an index over the number of estimated coefficients \textit{(`supports')} for attribute $k$,  $\omega_{n,k,q}$ gives the share of taste $q$ for attribute $k$ for individual $n$ and is itself based on a function of sociodemographic variables (with $\sum_{q=1}^{Q_k} \omega_{n,k,q} = 1,\forall k,n$) and $\Lambda_{k,m}$ is a matrix of combinations of coefficients (see the example in Figure \ref{fig:combinations}), that points at which coefficient $q$ to use for attribute $k$ under mixture $m$. Thus, for example, in the case where there are $K=6$ attributes and $Q_k=2 (\forall k)$ supports for each attribute, there will be a total of $M=2^6=64$ combinations of attribute sets (mixtures).

There are numerous methods for the incorporation of socioeconomic variables in a discrete mixture model. However, a complication arises due to the fact that there are so many different mixtures (64 in the above example), thus an approach such as following Equation \ref{eq:LC_socios} quickly becomes computationally infeasible. To reduce the complexity of the model, instead the impact of each sociodemographic variable can be considered for each attribute, resulting in K sets of coefficients rather than M. Then, since there are $Q_k$ supports for each attribute, we have weights for coefficient $q^*$ calculated by:  
\begin{equation}\label{eq:DM_Socios}
    \omega_{n,\textcolor{red}{k,q^*}} = \frac{\exp(\Delta_{\textcolor{red}{k,q^*}} + \sum_{l=1}^{L}{\left(z_{n,l} \cdot \zeta_{l,\textcolor{red}{k,q^*}})\right)}}{\textcolor{red}{\sum_{q=1}^{Q_k}}\exp(\Delta_{\textcolor{red}{k,q}} + \sum_{l=1}^{L}{\left(z_{n,l} \cdot \zeta_{l,\textcolor{red}{k,q}})\right)}},
\end{equation}
where the only difference between this equation and Eq. \ref{eq:LC_socios} for LC models is the different indices, highlighted in red, which means that we now have estimated coefficients $\zeta_{l,k,q}$ and $\Delta_{k,q}$ that vary over attributes and coefficients, rather than over classes. In the above example with 2 coefficients for 6 classes, fixing $\Delta_{k,1}=0$ and $\zeta_{l,k,1}=0$ for identification, results in $6$ sets of estimated sociodemographic interactions (one for each attribute) rather than one to distinguish the allocation between $2$ classes in a latent class model. This, by definition, means that sociodemographics are associated with specific differences in tastes (e.g. younger individuals are more cost sensitive) rather than a set of tastes as is the case in standard latent class models (e.g. younger individuals are more cost sensitive and less quality sensitive, etc.). 

A notable difference in the properties of discrete mixture models is that the preferences for different attributes have zero correlation if sociodemographic variables are fixed to zero (i.e. $\zeta_{l,k,q}=0, \forall l,k,q$). This is a result of weights $\omega_{n,k,q}$ being conditionally independent across $k$, i.e. the likelihood of having preference for attribute $A$ of $\beta_{A_1}$ is independent of the likelihood for having preference $\beta_{B_1}$ for attribute $B$.

\subsection{Generalised discrete mixture models}

Previous research has consistently demonstrated that a correlated mixed logit model outperform mixed logit models that assume independent random distributions for some parameters \citep{hess2017correlation,wang2021correlated} including in health applications \citep{mott2021valuing}. Thus, whilst reducing the level of correlation between tastes for different attributes may be desired, zero correlation (as is the case under discrete mixtures) may be a poor representation of the underlying relationship between tastes for different attributes. We propose two methods to account for the introduction of some correlation between tastes. 

The first involves `grouping' of certain attributes. For example, if we were to estimate coefficients associated with the different attribute levels (e.g. a coefficient for a `low', `medium' or `high' chance of side effects), then it is reasonable to expect some level of correlation in these tastes (e.g. a decision-maker who is very sensitive to a medium chance of side effects should also be very sensitive to a high chance of side effects). In such cases, a GDM would still estimate $Q_k$ different coefficients (supports) for the different attributes, but we might have, e.g. $\beta_{A_1}$ always in mixtures with $\beta_{B_1}$ but in mixtures with $\beta_{C_1}$ and $\beta_{C_2}$. The total number of mixtures would thus depend on the number of groupings of attributes, where grouping all attributes together corresponds to a latent class model. Mathematically, this would be represented by replacing the attribute $k$ with the attribute group $g$ ($1 \leq G \leq K)$ in Equations \ref{eq:DM_shares} and \ref{eq:DM_Socios}, where each mixture containing element $q$ of $g$ would contain element $q^{th}$ element of each attribute $k$ within group $g$. The total number of groups, $G$, varies between $1$ (corresponding to collapsing back to a latent class model) and $K$ (corresponding to collapsing back to the standard discrete mixture model). A disadvantage of this approach is that the analyst has to predefine the attribute groupings.

The second option for generalisation allows for additional correlation between preferences for different attributes by instead estimating additional boosts $\Delta_{r}$ that apply to a given set of mixtures $(m \in \mathcal{S}_r)$. To include this in the model, first note that the construction of Eq. \ref{eq:DM_shares} ensures that $\sum_{m=1}^{M}\pi_{m,n}=1, \forall n$. Thus rewriting Eq. \ref{eq:DM_shares} with a denominator, we have the share of mixture $m^*$ defined:

\begin{equation}\label{eq:DM_newshares}
    \pi_{m^*,n} = \frac{\prod_{k=1}^{K} \left( \sum_{q=1}^{Q_k} \omega_{n,k,q} \cdot (\Lambda_{k,m^*} = q) \right)}{\sum_{m=1}^{M}\left(\prod_{k=1}^{K} \left( \sum_{q=1}^{Q_k} \omega_{n,k,q} \cdot (\Lambda_{k,m} = q) \right)\right)}.
\end{equation}

We now introduce the boosts $\Delta_r$:
\begin{equation}\label{eq:GDM_shares}
    \pi_{m^*,n} = \frac{\exp\left(\sum_{r=1}^R \Delta_r \cdot (m^* \in \mathcal{S}_r)+ \ln\left(\prod_{k=1}^{K} \left( \sum_{q=1}^{Q_k} \omega_{n,k,q} \cdot (\Lambda_{k,m^*} = q) \right)\right)\right)}{\sum_{m=1}^{M} \exp\left(\sum_{r=1}^R \Delta_r \cdot (m \in \mathcal{S}_r)+ \ln\left(\prod_{k=1}^{K} \left( \sum_{q=1}^{Q_k} \omega_{n,k,q} \cdot (\Lambda_{k,m} = q) \right)\right)\right)},
\end{equation}

\noindent where a boost $\Delta_r$ is applied if mixture $m^*$ is in the subset $\mathcal{S}_r$. Note that if $\Delta_r=0, \forall r$, then Eq. \ref{eq:GDM_shares} collapses back to Eq. \ref{eq:DM_newshares}, which is equivalent to \ref{eq:DM_shares}, meaning we obtain the standard discrete mixture model. As a contrast, if we specify boosts $\Delta_r$ for classes containing the $q^{th}$ coefficient for each attribute when $Q_k=Q, \forall k,$ (e.g. when $Q_k=2, \forall k,$ and we have a shift for the classes represented by $m=1$ and $m=64$ in Figure \ref{fig:combinations}), high estimates for these boosts correspond to the model collapsing towards a latent class model, for which there are no classes with shared coefficients. For $0<\Delta_r<\infty$, then, the GDM will improve on DM and LC models. 

\section{Visual overview of models}
\begin{figure}[ht]
\centering\includegraphics[width=0.8\linewidth]{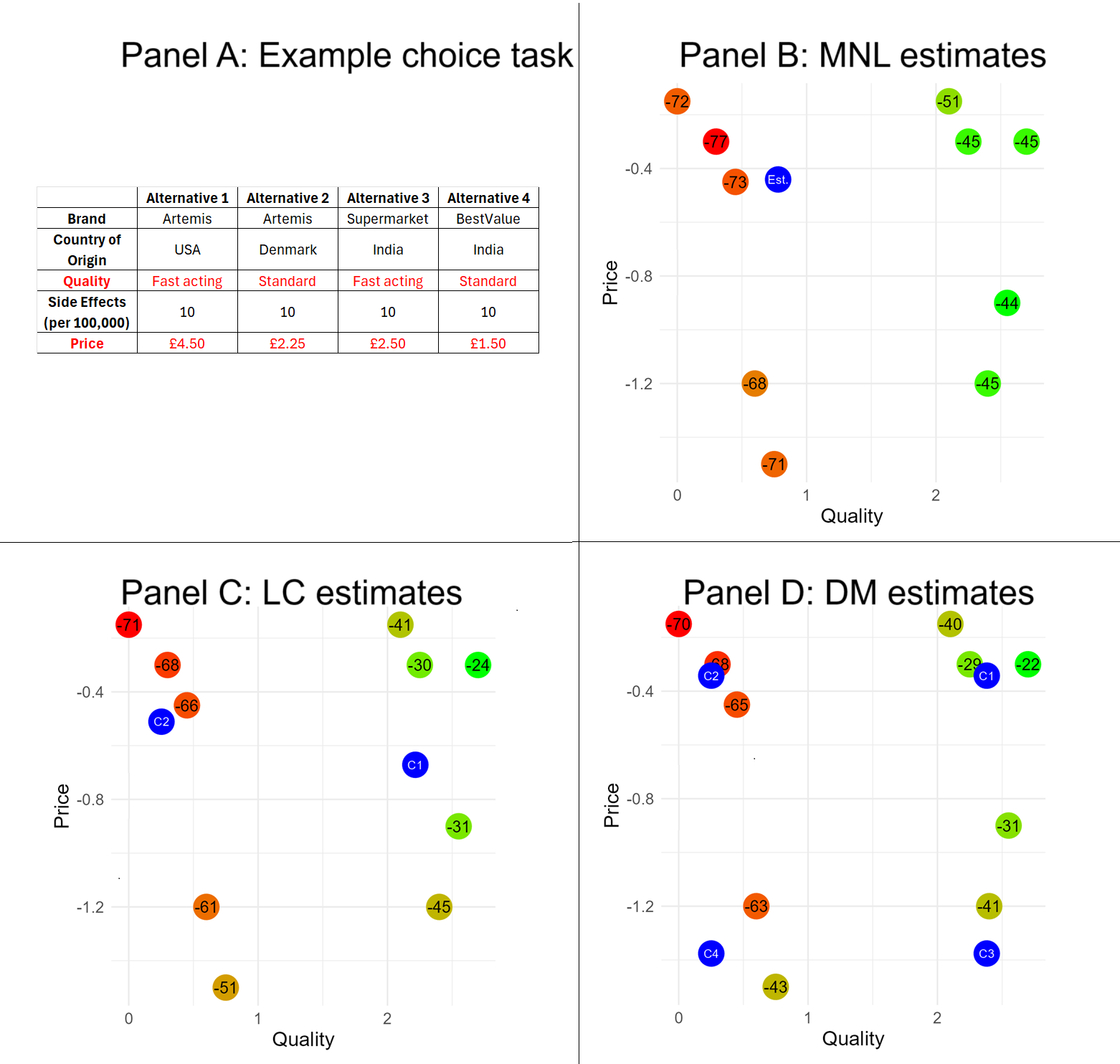}
    \caption{Panel A gives an example choice task. Panels B-D show the model estimates (represented by blue circles) for MNL, LC and DM models, respectively. Each other circle represents the log-likelihood contribution for explaining choices made by the different simulated patients, with the location of these circles representing the underlying preference of these patients. The circles are coloured green to red depending on their relative contribution to the log-lieklihood.}
    \label{fig:ModelVisuals}
\end{figure}

To demonstrate how the different models capture preferences, we build on the example shown in Figure \ref{fig:PrefVisual}, where 10 simulated patients have different tastes for the quality and costs of the alternatives. These tastes were used to generate 50 choices per patient in choice tasks based on the simulated `drug choice' dataset given on the \href{https://www.apollochoicemodelling.com}{Apollo} choice modelling website. An example choice task is given in Panel A of Figure \ref{fig:ModelVisuals}.

\noindent For simplicity, we assume that patients only care about the price and quality of the alternatives. Under a \textbf{MNL model}, only one coefficient is estimated for the sensitivity to price, and one for the sensitivity to quality (the blue marker denoting both the price parameter of around -0.45, and the quality parameter of around 0.9). Panel B shows that these estimates do not capture average preferences, but are instead skewed towards the preferences of the less sensitive individuals. This is a result of the higher contribution to utility from the `harder to explain' choices made by these individuals (whose choices are by definition more random). Consequently, the MNL model is improved by better explaining these individuals with respect to the more cost and quality sensitive individuals. This is in line with previous results that demonstrate that failing to account for preference heterogeniety can lead to bias in parameter estimates \citep{van2010biases}. If  we use a \textbf{latent class model}, we allow for two sets of preferences, $\beta_{Q_1}$ \& $\beta_{P_1}$, and $\beta_{Q_2}$ \& $\beta_{P_2}$. Panel C in Figure \ref{fig:ModelVisuals} shows the estimates for these for the LC model. Notably, the model appears to capture quality sensitivity, with the estimates $\beta_{Q_1}$ and $\beta_{Q_2}$ very different (approximately 0.25 and 2.25, respectively), but not price sensitivity, with the estimates $\beta_{P_1}$ and $\beta_{P_2}$ more similar (approximately -0.5 and -0.7, respectively). Finally, Panel D shows the estimates for the \textbf{discrete mixture model}. It specifically allows for combinations of preferences, resulting in 4 classes of preferences represented by a `grid' of the coefficient estimates. This substantially reduces the log-likelihood contribution of the most cost sensitive individuals with respect to the LC model results. Notably, in the context studied here, the correlation between price sensitivity and quality sensitivity was low ($-0.0037$). In the subsequent section on applications of the different models to simulated data, we explore how the models perform with different amounts of correlation between the sensitivities.

\section{Empirical application: simulated data}
We now test different simulated datasets, where we assume `latent class' and `discrete mixture' type preferences, respectively, across different datasets. The aim of the simulation exercise is to test the different model structures with respect to the underlying preference structure. We now detail the generation of the simulated dataset, and the model results, in turn.

\subsection{Generation of simulated data}

\begin{figure}[ht]
    \centering
    \includegraphics[width=1\linewidth]{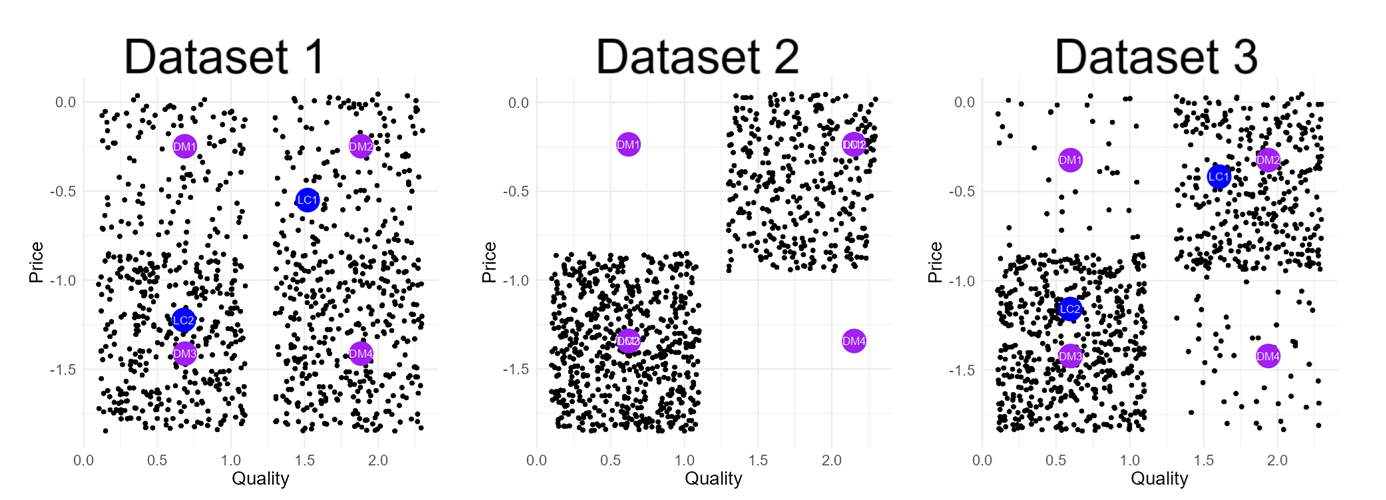}
    \caption{The simulated preference distribution across three datasets. The first assumes individuals are equally likely to be in each `quadrant'. The second specifically assumes preferences are correlated, and that individuals are either (but not both) cost sensitive or price sensitive. The third assumes a strong level of correlation whilst allowing for some individuals (10\%) to exhibit either insensitive or sensitive tastes to both quality and price.}
    \label{fig:Sim}
\end{figure}

Using the same Apollo `drug choice' dataset as before, we simulate three new datasets. The preferences for quality and price vary across 1,000 simulated respondents, where the level of correlation between these respondents` quality and price sensitivities differ across the datasets. This allows us to further test the relative difference in capturing preferences between DM and LC models. In the first case, 50\% of respondents are assumed to be either cost sensitive or quality sensitive. In the second, 100\% are, and in the final dataset, 90\% are. Thus, in theory, the DM model should perform best for the first dataset, which has preferences equally distributed across all four quadrants (see Figure \ref{fig:Sim}). For the second dataset, we have all preferences in either the quadrant representing high price sensitivity and low quality sensitivity, or vice versa. The LC should perform relatively well here. In the final dataset, a purposefully limited number of individuals are in the off-diagonal quadrants representing sensitivity to both attributes or neither. In this case, the DM should be preferred to LC, but to  lesser degree that in dataset 1. 

\subsection{Model results}
The log-likelihoods for the discrete mixture and latent class models are given in Table \ref{tab:Sim}. 

\begin{table}[ht]
    \centering
    \begin{tabular}{|l|l|l|l|}
    \hline
        & Dataset 1 & Dataset 2 & Dataset 3 \\ \hline
        Cost/quality sensitive & 25\% & 0\% & 5\% \\ \hline
        Cost/quality insensitive & 25\% & 0\% & 5\% \\ \hline
        Cost sensitive, quality insensitive & 25\% & 50\% & 45\% \\ \hline
        Cost insensitive, quality sensitive & 25\% & 50\% & 45\% \\ \hline
        Cost/quality correlation & 0.05 & 0.74 & 0.59 \\ \hline
        Discrete mixture log-likelihood & -10,073 & -10,102 & -10,111 \\ \hline
        Latent class log-likelihood & -10,264 & -10,060 & -10,152 \\ \hline
        Difference in LL & 191 & -42 & 41 \\ \hline
        Generalised discrete mixture log-likelihood & 10,058 & 10,042 & 9,996 \\ \hline
    \end{tabular}
    \caption{Simulated data settings and results.}
    \label{tab:Sim}%
\end{table}

In dataset 1 and 2, results are in line with expectations, with a DM model performing much better for dataset 1, and LC model performing better for dataset 2. This is represented visually by the model estimates in Figure \ref{fig:Sim}, where purple circles represent the preferences of the four different classes in the DM model, and blue circles represent the two different classes in the LC model. The LC model fails to cover all types of preferences in dataset 1, whilst the DM model has classes that do not closely represent any respondent preferences in dataset 2. Interestingly, the DM model comfortably outperforms the LC model for dataset 3, despite there only being a few individuals in the off-diagonal quadrants.
\section{Empirical application to SP data}
\subsection{Dataset I}
\begin{figure}[ht!]
    \centering
    \includegraphics[scale=1]{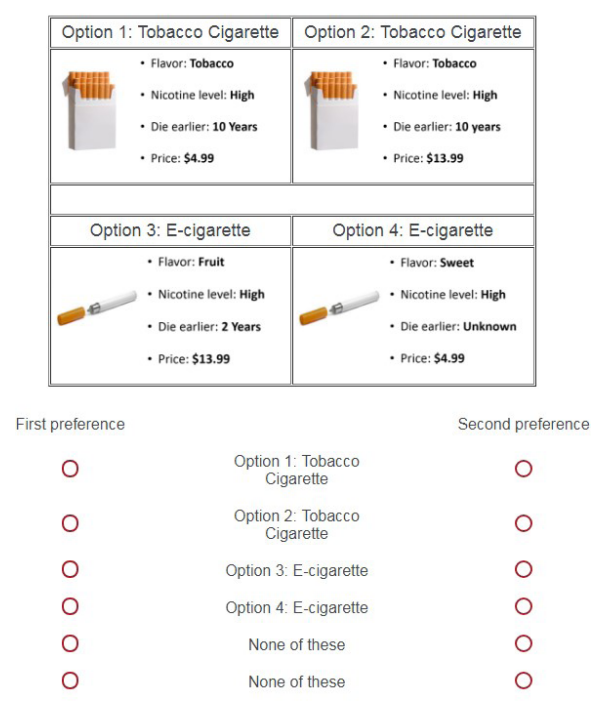}
    \caption{An example choice task from dataset 1 \citep{buckell2019should}}
    \label{fig:data}
\end{figure}

We first test the different models on a dataset from a discrete choice experiment on tobacco preferences.  2,031 smokers from the US complete 12 choice tasks each. Each choice task had two cigarette options, two e-cigarette options and an opt-out \citep{buckell2019should}. The cigarettes were described by flavour, cost, nicotine amount and `life years lost’, with an example choice task being displayed in Figure \ref{fig:data}.

Previous models fitted to this data \citep{buckell2019should} found extensive deterministic heterogeneity with e.g. African Americans and more educated individuals more likely to choose menthol cigarettes over tobacco cigarettes, and younger adults more likely than older adults to choose tobacco, menthol or flavoured e-cigarettes. \citet{buckell2021kicking} further demonstrated that there is substantial heterogeneity in preferences for nicotine levels, with addicted smokers valuing nicotine much more than less addicted smokers. Finally, \citet{meester2023can} found evidence of substantial taste heterogeneity and also decision-rule heterogeneity, with latent class models with different model structures in the respective classes performing better than models with the same model structure across classes. 

To test the different model structures of interest here, we directly build upon the models developed by \citep{buckell2019should}. We further assume that there is unobserved heterogeneity for the preference towards different types of cigarettes (tobacco or menthol), e-cigarettes (tobacco, menthol or flavoured) or the opt-out. In this case, a latent class model would assume preferences for different types of cigarettes or e-cigarettes are correlated, whereas a discrete mixture model would not. This may be beneficial for a case where preferences for opting out are uncorrelated with preferences towards cigarettes or e-cigarettes. A GDM allows for a combination, meaning a model could e.g. capture correlation across e-cigarette options whilst recognising the lack of correlation with the opt out.

\subsection{Model specification and estimation I}\label{sec:Spec}

To estimate preferences for different flavoured products, we interacted the `product constant terms' (e-cigarette/cigarette, see Figure \ref{fig:data}) 
with each of the flavours (tobacco/menthol/flavoured) to examine the product–flavour pairings directly. 
The utility for alternative $i$ for individual $n$ in choice task $t$ is therefore estimated:
        \begin{linenomath*}               \begin{equation}\label{eq:utility1}
                \begin{split}
                    V_{nti} &= \delta_{cig-tobacco} \cdot cig_{nti} \cdot tobacco_{nti}\\
                    &+ \delta_{cig-menthol} \cdot cig_{nti} \cdot menthol_{nti}\\
                    &+ \delta_{ecig-tobacco} \cdot ecig_{nti} \cdot tobacco_{nti} \\ &+ \delta_{ecig-menthol} \cdot ecig_{nti} \cdot menthol_{nti} \\ &+ \delta_{ecig-fruitsweet} \cdot ecig_{nti} \cdot fruitsweet_{nti} \\&+ \delta_{optout} \cdot optout_{nti}\\ 
                    &+ \beta_{price} \cdot price_{nti}\\
                    &+ \beta_{nonic} \cdot nonic_{nti} + \beta_{lownic} \cdot lownic_{nti} + \beta_{mednic} \cdot mednic_{nti} + \beta_{highnic} \cdot highnic_{nti} \\
                    &+ \beta_{unknown} \cdot unknown_{nti} + \beta_{2yr} \cdot 2yr_{nti} + \beta_{5yr} \cdot 5yr_{nti} + \beta_{10yr} \cdot 10yr_{nti},
                \end{split}
            \end{equation}
        \end{linenomath*}
where $\delta$ and $\beta$ are estimated coefficients, and all attribute variables except price are binary such that there is a coefficient associated with each attribute level.\footnote{Note that $\delta_{cig-tobacco}$,$\beta_{mednic}$, and $\beta_{10yr}$ are fixed to a value of 0 for identification purposes.}

We then test four models with various means to capture unobserved heterogeneity:
\begin{enumerate}
    \item A base multinomial logit model (with no means to capture heterogeneity), based on Eq. \ref{eq:utility1}.
    \item A latent class model with two classes with different alternative specific constants ($\delta$) for the different types of cigarettes/e-cigarettes.
    \item A discrete mixture model estimating two constants per product-flavour pairing, but with all possible combinations of coefficients. This results in a total of $2^5=32$ classes.
    \item A generalised discrete mixture model that captures the correlation between preferences for different types of e-cigarettes (to try and capture the fact that an individual who likes one type of e-cigarette may also like another type).
\end{enumerate}

Two different versions of each of the four models are tested. The first set do not include sociodemographic variables thus do not capture deterministic heterogeneity, whereas the second set does. Results from previous research using this dataset \citep{buckell2019should,buckell2021kicking,meester2023can} are used to select the following set of binary variables. These are whether the participant is:

\begin{enumerate}
    \item $\zeta_{n,young}$, younger (24 years old or less)
    \item $\zeta_{n,older}$, older (55 years old or more)
    \item $\zeta_{n,fem}$, female
    \item $\zeta_{n,afram}$, African American
    \item $\zeta_{n,asian}$, Asian
    \item $\zeta_{n,hisp}$, Hispanic
    \item $\zeta_{n,other}$, of another (non-white) ethnicity
    \item $\zeta_{n,heduc}$, has higher education (BSc or equivalent or more)
    \item $\zeta_{n,hinc}$, a high earner (\$55k or more)
    \item $\zeta_{n,lhous}$, from a small household (2 or less individuals)
    \item $\zeta_{n,lowsr}$, of low self-reported health
    \item $\zeta_{n,vaper}$ a vaper
    \item $\zeta_{n,dual}$ a user of both cigarettes and e-cigarettes
    \item $\zeta_{n,tquit}$ an attempted quitter 
    \item $\zeta_{n,quit}$ a successful quitter.
\end{enumerate}

\subsection{Results I}
All 8 model results are given in Table \ref{tab:Results1}.

\begin{table}[!ht]
    \centering
    \begin{tabular}{|c|c|c|c|c|c|c|c|}
    \hline
        ~ & \multicolumn{3}{c|}{BASE MODEL} & \multicolumn{3}{c|}{DETERMINISTIC HETEROGENEITY} & ~ \\  \hline
        ~ & free pars & LL & BIC & free pars & LL & BIC & LL diff  \\ \hline
        MNL & 12 & -37,200.50 & 74,522 & 87 & -35,527.72 & 71,934 & 1,673  \\ \hline
        LC & 18 & -33,322.52 & 66,827 & 33 & -33,216.33 & 66,766 & 106  \\ \hline
        DM & 22 & -29,715.55 & 59,653 & 97 & -29,303.85 & 59,588 & 412  \\ \hline
        GDM & 24 & -28,963.57 & 58,170 & 99 & -28,664.97 & 58,330 & 299  \\ \hline
    \end{tabular}
    \caption{Model results from multinomial logit (MNL), latent class models (LC), discrete mixture models (DM) and generalised discrete mixture models (GDM), with and without deterministic heterogeneity. LL(0)=-39,225.22.}
    \label{tab:Results1}
\end{table}

Key results from these models can be summarised as follows:
\begin{enumerate}
    \item There is substantial random and deterministic preference heterogeneity. Both models that allow for random preferences (base LC, DM, GDM models) and the MNL model allowing for observed heterogeneity outperform the base MNL model.
    \item Discrete mixture models substantially outperformed the latent class model counterparts, even after accounting for their greater complexity through consideration of BIC values.
    \item Generalised discrete mixture models further improved model performance, indicating correlation between preferences for the different flavours of e-cigarettes.
    \item Accounting for random heterogeneity had a much greater impact on model performance than accounting for deterministic heterogeneity. However, it should be noted that this does not imply models should forgo the inclusion of variables to account for observed indiviudal differences. 
    \item Specifically, DM models were far better able to account for observed differences in preferences than LC models. This suggests that subgroups of individuals have preferences for specific product-flavour pairs, rather than for a set of preferences (as in LC models). 
    \item The overall best performing model was GDM with deterministic heterogeneity, but was the base GDM model if accounting for model complexity.
    
\end{enumerate}

\subsection{Dataset II}

\begin{figure}[ht!]
    \centering
    \includegraphics[scale=1]{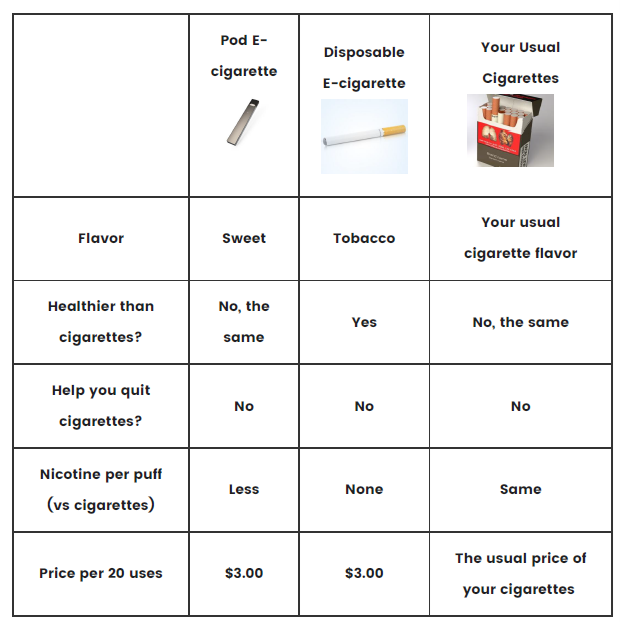}
    \caption{An example choice task from dataset 2 \citep{Buckelle173}}
    \label{fig:dataII}
\end{figure}

The second empirical dataset is taken from \citep{Buckelle173}. Here, 2,000 smokers in the US who had little or no interest in quitting (NIQ) completed 12 choice tasks. Tasks offered pod and disposable e-cigarettes, and the choice of "your usual cigarette". Alternatives were described by: flavours, whether the product is healthier than cigarettes, whether the product help to quit cigarettes, nicotine, and prices. An example choice task is given in Fig. \ref{fig:dataII}. 

Previous analyses \citep{Buckelle173} used a hybrid, latent class model with 2 classes, indicating heterogeneity across alternatives and attributes. A set of demographic variables, e-cigarette use, and a health behaviours latent variable explained class membership.  For simplicity and exposition, we forego the latent variable here. Instead, we focus on the comparison of latent classes, DM, and GDM in this setting, with and without deterministic heterogeneity using demographic variables. 

\subsection{Model specification II}

The deterministic component of utility is described by product constant terms, 

        \begin{linenomath*}               \begin{equation}\label{Eq:LondonCycle}
                \begin{split}
                    V_{nti} &= \delta_{cigarette} \cdot cigarette_{nti}\\
                    &+ \delta_{pod-ecig} \cdot pod_{nti}\\
                    &+ \delta_{disposable-ecig} \cdot disposable_{nti}\\
                    &+ \delta_{menthol-smoker} \cdot menthol-smoker_{nti}\\
                    &+ \beta_{tobacco} \cdot tobacco_{nti} + \beta_{menthol} \cdot menthol_{nti} + \beta_{fruit} \cdot fruit_{nti}+ \beta_{sweet} \cdot sweet_{nti}\\                    
                    &+ \beta_{healthier-than-cigarettes} \cdot healthier-than-cigarettes_{nti}\\
                    &+ \beta_{helps-you-quit} \cdot helps-you-quit_{nti}\\
                    &+ \beta_{no-nicotine} \cdot no-nicotine_{nti} + \beta_{less-nicotine} \cdot less-nicotine_{nti}  \\
                    & + \beta_{same-nicotine} \cdot same-nicotine_{nti} + \beta_{more-nicotine} \cdot more-nicotine_{nti}\\
                    &+ \beta_{cheaper} \cdot cheaper_{nti} + \beta_{expensive} \cdot expensive_{nti}
                \end{split}
            \end{equation}
        \end{linenomath*}
where $\delta$ and $\beta$ are estimated coefficients, and all attribute variables except prices are binary such that there is a coefficient associated with each attribute level. Prices are modelled relative to the respondent's current cigarette prices. In some cases the e-cigarette alternative is more expensive, in other cases cheaper, and in some cases the same price. $\beta_{cheaper} $ and $\beta_{expensive} $ measure preferences for price differentials. \footnote{Note that $\delta_{cigarette}$,$\beta_{tobacco}$ and $\beta_{same-nicotine}$ are fixed to a value of 0 for identification.}

We use a base MNL model along with 2- and 3-latent class models (NB - in preliminary testing, a 4-latent class model did not estimate on this data). To each LC model, we add deterministic heterogeneity (`Dethet'). We thus have a set of 5 models (MNL - (i); 2-LC (ii); 2-LC-Dethet (iii); 3-LC (iv); 3-LC-Dethet (v)) against which to compare DM and GDM models. (NB - at the time of writing, we are part way through our modelling process. We show superior performance to base models, but are yet to optimise specifications of these models.)

\subsection{Results II}

\begin{table}[ht]
\small
    \centering
    \begin{tabular}{|c|l|c|c|c|c|c|}
        \hline
        \textbf{Section} & \textbf{Model} & \textbf{pars} & \textbf{supports} & \textbf{groups} & \textbf{classes/mixtures} & \textbf{BIC} \\
        \hline
        i & MNL & 13 & 1 & 1 & 1 & 31,216.58 \\
        \hline
        \multirow{9}{*}{ii} & 2-Latent Class MNL & 27 & 2 & 1 & 2 & 22,767.78 \\
        & DM 2 pars 2 supports & 17 & 2 & 2 & 4 & 22,574.40 \\
        & DM 3 pars 2 supports & 19 & 2 & 3 & 8 & 22,455.80 \\
        & DM 4 pars 2 supports & 21 & 2 & 4 & 16 & 22,657.73 \\
        & DM 5 pars 2 supports & 23 & 2 & 5 & 32 & 22,479.14 \\
        & DM 7 pars 2 supports & 26 & 2 & 7 & 128 & 21,858.17 \\
        & DM 8 pars 2 supports & 28 & 2 & 8 & 256 & 21,553.91 \\
        & DM 13 pars 2 supports 5 groups & 31 & 2 & 5 & 32 & 21,935.70 \\
        \hline
        \multirow{3}{*}{iii} & 2-Latent Class MNL Dethet & 36 & 2 & 1 & 2 & 22,610.45 \\
        & DM 2 pars 2 supports Dethet & 35 & 2 & 2 & 4 & 22,384.23 \\
        & DM 3 pars 2 supports Dethet & 43 & 2 & 2 & 8 & 22,233.71 \\
        \hline
        \multirow{3}{*}{iv} & 3-Latent Class MNL & 40 & 3 & 1 & 3 & 21,737.41 \\
        & DM 2 pars 3 supports & 21 & 3 & 2 & 9 & 21,831.40 \\
        & GDM 2 pars 3 supports & 23 & 3 & 2 & 9 & 21,223.75 \\
        \hline
        \multirow{2}{*}{v} & 3-Latent Class MNL Dethet & 55 & 3 & 1 & 3 & 21,548.48 \\
        & DM 2 pars 3 supports Dethet & 57 & 3 & 2 & 9 & 21,235.05 \\
        \hline
    \end{tabular}
    \caption{Model Comparison Table from Dataset 2}
    \label{tab:model_comparison}
\end{table}

 Table \ref{tab:model_comparison} shows the results from Dataset II. Comaprator models (ii)-(v) are compared to DM and/or GDM models. For model (ii), there are 27 parameters and 2 classes. We then estimated a series of DM models, each increasingly allowing for more discrete mixtures. The number of "pars" is the number of parameters over which mixing is specified. The number of "supports" is the number of supports for that parameter in the mixture. The model "DM 2 pars 2 supports" mixes ASCs for pod and disposable e-cigarettes - i.e., two parameters - with two supports - i.e. a preference A and B for both ASCs. Hence, there are $2^2 = 4$ possible combinations of preferences, and so 4 mixtures. We subsequently add in mixtures for more parameters in the model. "DM 8 pars 2 supports" has 8 parameters with two supports and therefore $2^8 = 256$ possible combinations of preferences and therefore 256 mixtures.  As in the MNL model, there are 13 parameters in the base model. To apply the DM to the full model would require $2^1{}^3 = 8,192$ possible preference combinations. It was not possible to estimate such a model. Instead, parameters were grouped into those similar to each other (e.g. a group of ASCs containing  parameters for pod and disposable e-cigarettes; and a group of flavor parameters). The DMs were applied over the groups, resulting in a more tractable model. "DM 13 pars 2 supports 5 groups" allows attribute-group-wise heterogeneity and to treat all parameters as discrete mixtures. In this model, then, we have $2^5=32$ combinations. This approach balances flexibility with tractability. With respect to fit, all models in (ii) have superior BICs to model (ii).   

 We next compare the 2-class LCMNL with deterministic heterogeneity (i.e., model (iii)) to DM models in section (iii). Here, superior BICs are achieved for a DM model using mixtures on two (of 13) parameters. Further gains are yielded with the DM treatment of a third parameter. 

 In section (iv), the 3-class LCMNL outperfoms the 2-parameter, 3-support DM model. However, we note estimation issues for this model: we reached a saddle point and were unable to find a superior solution. In this case, the GDM version of the same model achieved a superior BIC to model (iv). In section (v), the 2-parameter, 3-support DM model with deterministic heterogeneity has a better BIC than the 3-class LCMNL with deterministic heterogeneity. 
 
 In cases (ii) to (v), with mixtures on only two (of 13) parameters, DM/GDM models outperform latent class counterparts. 

\section{Conclusions and next steps}
In this paper, we proposed discrete mixture models as an alternative to latent classes and mixed multinomial logit models for capturing random preference heterogeneity. We developed a generalised discrete mixture model which will perform at least as well as a discrete mixture or latent class, if not better. 

The structure of latent class models means that by definition, some level of correlation between preferences for different attributes is imposed. In contrast, a discrete mixture model as defined by \citet{hess2007systematic} assumes zero correlation between the preferences for different attributes, as a result of the probability of having different tastes for one attribute being independent of the probability for having a specific taste for another attribute. Our new generalised discrete mixtures that either allow for groups of attributes to be contained together (introducing correlation only where desired) or for some mixtures with specific combinations of preferences to occur more often (allowing for correlation dependent on empirical results). Both versions are generalisations that can collapse to standard discrete mixtures or latent class models. As the latter form of GDM is estimated, it by definition performs at least as well as the better of an LC and a DM model. 

We find that discrete mixture models substantially outperform latent class models in all tested cases using two smoking preference datasets. Furthermore, GDMs add further improvements and DMs better capture deterministic heterogeneity than LCs. This implies that whilst there are some correlations in preferences, the greater flexibility of the new models allows for a better representation of preferences.

Exercises on simulated data indicate the strict and improbable conditions under which a LC model would be preferred to a DM. This appears to be corroborated by the empirical data in two cases. The common practice of labelling latent classes, therefore, we believe is questionable. The GDM reanalysis of \citep{Buckelle173}'s data casts doubt on their characterisation of NIQ smokers as "switchers" and "non-switchers". The preference structure appears vastly more complex than this, and such an interpretation of behaviours could mislead if taken at face value. If such a grouping is reasonable on the data, the GDM would recover this. 

 Next steps include comparisons with continuous mixture models (multinomial mixed logit models) and an investigation of GDM properties such as IIA, and model outputs such as willingness-to-pay and elasticities. 

We believe that the GDM can be a useful tool in modelling preference heterogenetiy in health choices.





\small
\bibliographystyle{apalike}
\bibliography{references}

\end{document}